# AI-Generated Images for representing Individuals

## Navigating the Thin Line Between Care and Bias

Julia C. Ahrend
Muthesius University
of Fine Arts & Design /
KielSCN, Germany
jahrend@kielscn.de

Björn Döge
Muthesius University
of Fine Arts & Design /
KielSCN, Germany
bdoege@kielscn.de

Tom M Duscher
Muthesius University
of Fine Arts & Design /
KielSCN, Germany
tduscher@kielscn.de

Dario Rodighiero
University of Groningen, Netherlands
d.rodighiero@rug.nl

## Abstract
This research discusses the figurative tensions that arise when using portraits to represent individuals behind a dataset. In the broader effort to communicate European data related to depression, the Kiel Science Communication Network (KielSCN) team attempted to engage a wider audience by combining interactive data graphics with AI-generated images of people. This article examines the project's decisions and results, reflecting on the reaction from the audience when information design incorporates figurative representations of individuals within the data.

## Introduction
Using generative AI to visualize parts of a statistical dataset highlights the tension between technical innovation and ethical responsibility. This is especially true when representing individuals or members of a specific community. It is difficult to create images that reflect the full human diversity of the data [1].

For instance, models such as Midjourney 5.2 struggle to produce images that accurately depict individuals with distinct personal characteristics, such as older age and ethnicity beyond dominant Western visual norms.

These limitations are not purely algorithmic. Any act of visual representation, including those made by designers, carries aesthetic assumptions and care. This is particularly relevant in the context of a mental health project, because depression is still socially stigmatized and often depicted in dark, melancholic imagery.

To explore alternative approaches, the research team conducted a participatory workshop where participants generated their own image prompts to represent depression. The variety of results challenged prevailing visual norms and highlighted how closely representation, bias, and identification are intertwined. In this context, designing with care means not only refining prompts, but also critically engaging with the social meanings that images carry and how they shape public perception.

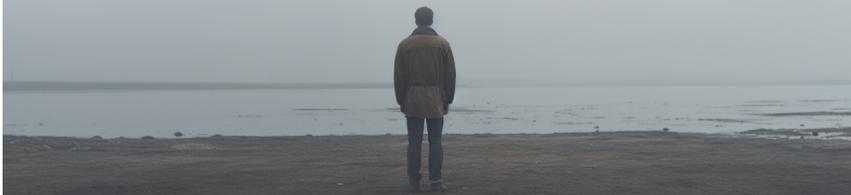

Fig. 1 (from [2]): Section of the article containing the interactive information graphic: In the stage displayed, the user has filtered for "male" and "35 to 44", the image has changed accordingly.

## Background of the Project

This project was part of an explorative design study, developed in collaboration between the KielSCN and the German science magazine Spektrum der Wissenschaft. The study examined how Midjourney could enable new approaches to visualizing statistical data on mental health, as depicted in [2, Fig. 1]. The resulting imagery was featured in an interview titled 'Depression. The Long Shadow of Society' [2].

The visualization draws on epidemiological data collected through the European Health Interview Survey (EHIS). This survey provides insights into the prevalence of depression across demographic groups in Europe [3]. Prevalence refers to the total number of cases of the disease in the population under consideration. Participants of the study indicated whether they had suffered from depression within the previous 12 months, or whether they experienced symptoms of the illness.

In terms of interaction, users can filter the dataset by age, gender, and country of origin. The background image adapts to these filters, allowing users to compare their demographic data with broader statistical patterns This personalization fosters an intimate and relatable engagement with the subject of depression, one of the central objectives of the project

In addition, the project aimed to enrich the data visualization with illustrative elements to convey the mood associated with depression by integrating images that represent individuals from the dataset. One underlying idea was that viewers may relate more strongly to personalized visualizations, potentially identifying figures reflecting their own demographic profile.

However, translating data into images introduced both technical and ethical challenges. As the project unfolded, it became clear that Midjourney's generative model 5.2 reproduces narrow aesthetic defaults and often struggles to depict diverse populations with nuance. This raised critical questions about what it means to care, represent, and generalize in the context of visual science communication.

## Interactivity:
## From Explanation to Exploration

In order to accommodate small screen sizes, the interface design is responsive. In the desktop version as illustrated in [2, Fig. 2], the bar chart itself functions as the filter mechanism, providing an integrated and exploratory interface. Users have the ability to directly interact with the bar charts in order to adjust parameters, thereby integrating the processes of data exploration and interaction. Conversely, the mobile version as shown in [2, Fig. 3] utilizes conventional drop-down filters, emphasizing usability and clarity on smaller screens where displaying the complete dataset is impractical. Here, emphasis is placed on percentage values that summarize the user's selected demographic configuration.

The project incorporates a second entry point in the form of a step-by-step introduction that leads users through the available filters and gradually enhances their comprehension. This structured interaction ends in the prompt to "filter data according to your characteristics," encouraging users to reflect on how their own identity relates to the data. The complete scaffold is illustrated as in [2, Fig. 4–7] on the subsequent page. These figures underscore the disparities in visual complexity between the desktop and mobile version.

While the project team has gathered preliminary insights using Google Analytics and heatmaps, a full evaluation has not yet been conducted. Such an evaluation would be necessary to better understand what specifically guided user engagement and how the interactive elements were experienced. However, the existing data showed that users did engage with the graph: one proportion proceeded through the step-by-step introduction. A comparison with other articles on Spektrum der Wissenschaft reveals that the interaction rate was above average (1 minute 57 seconds versus 43 seconds for comparable non-interactive articles), suggesting that the interactive format may have increased user engagement.

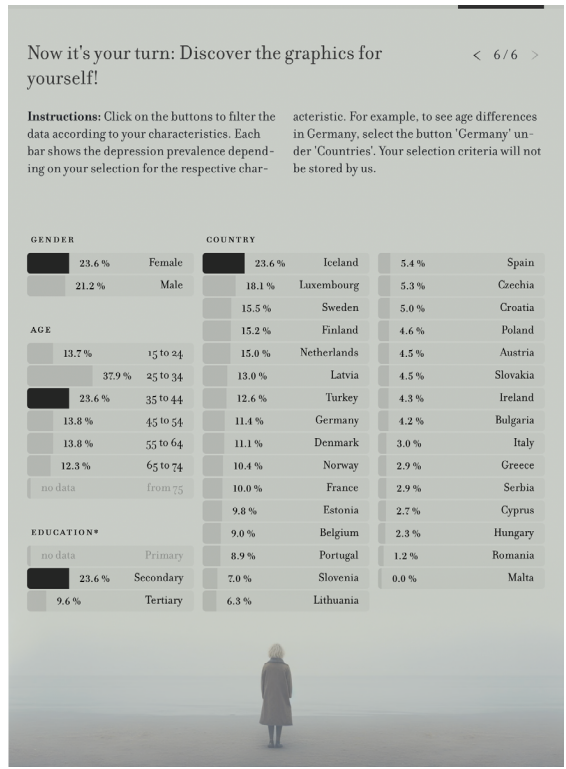

Fig. 2: Desktop version (from [2]).

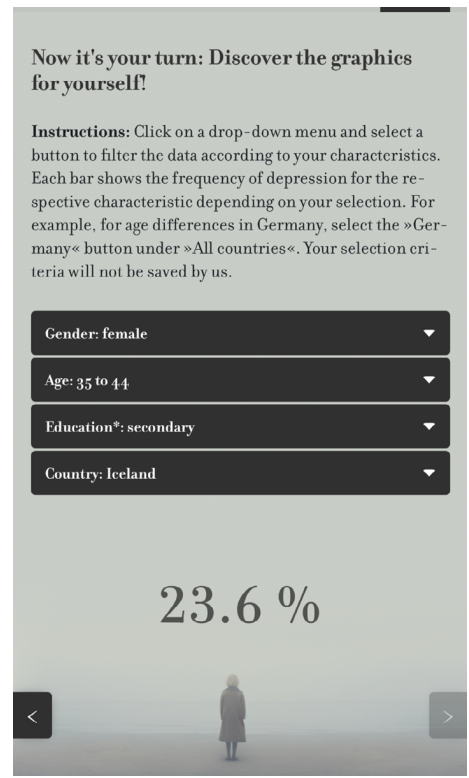

Fig. 3: Mobile Version (from [2]).

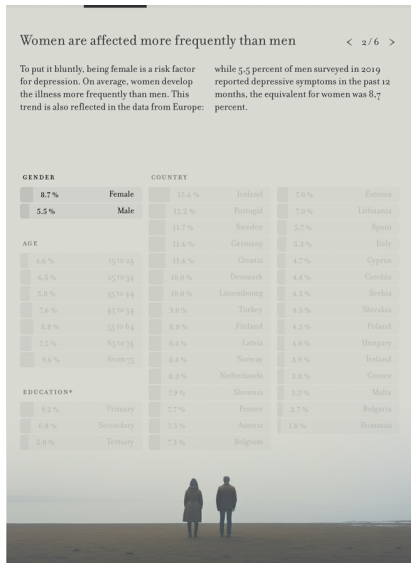
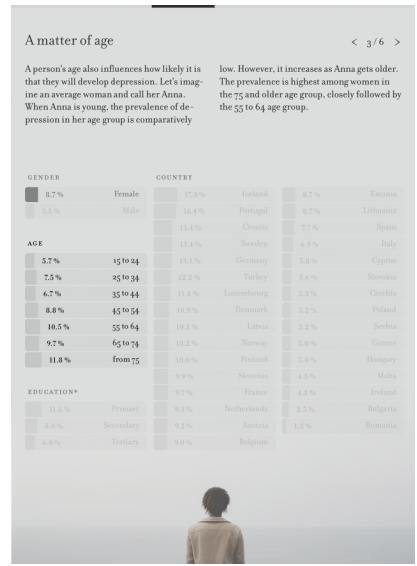
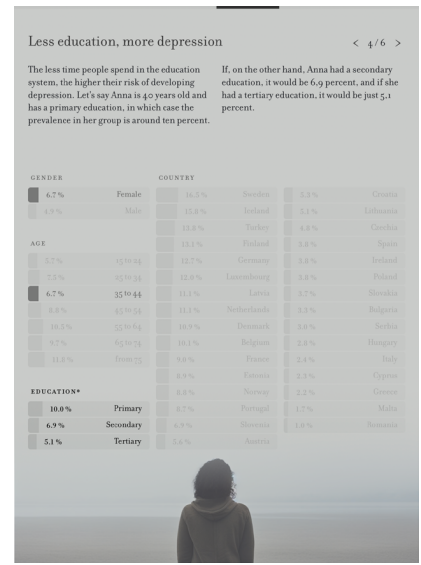
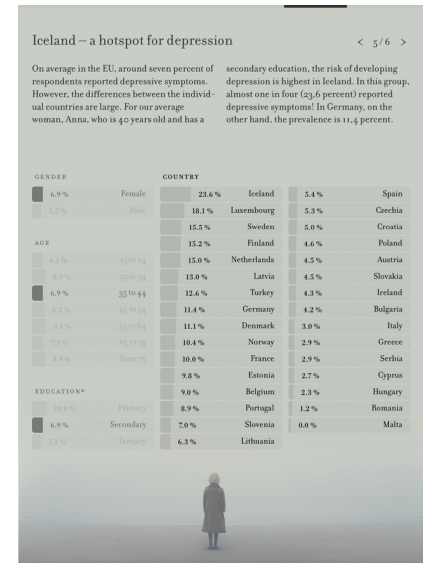

Top: Fig. 4a. Desktop;
Bottom: Fig. 4b. Mobile;
Screen Gender (from [2]).

Top: Fig. 5a. Desktop;
Bottom: Fig. 5b. Mobile;
Screen Age (from [2]).

Top: Fig. 6a. Desktop;
Bottom: Fig. 6b. Mobile;
Screen Education (from [2]);

Top: Fig. 7a. Desktop;
Bottom: Fig. 7b. Mobile;
Scaffold Country (from [2]).

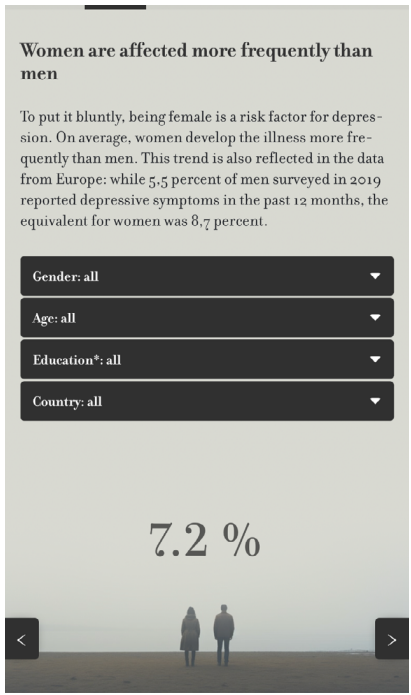
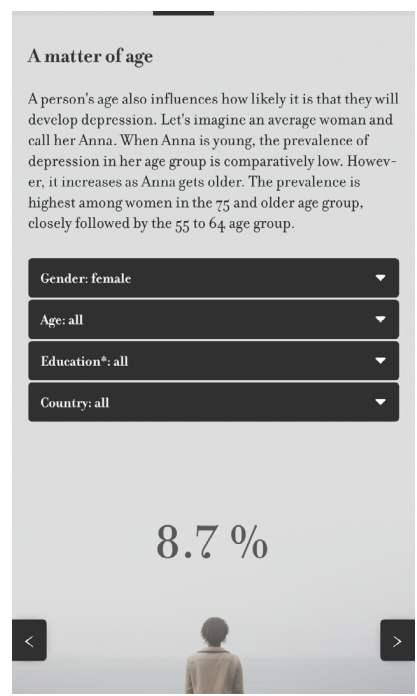
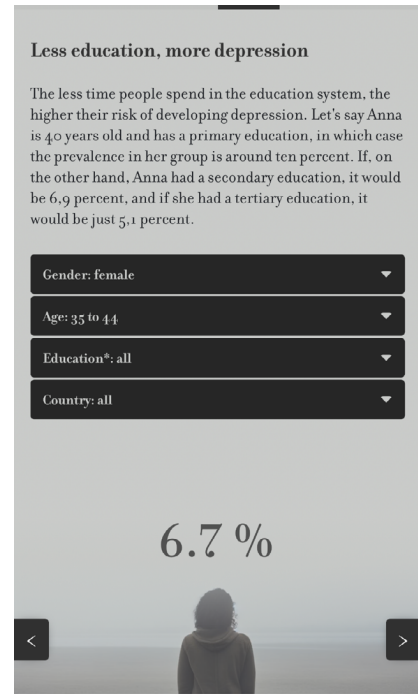
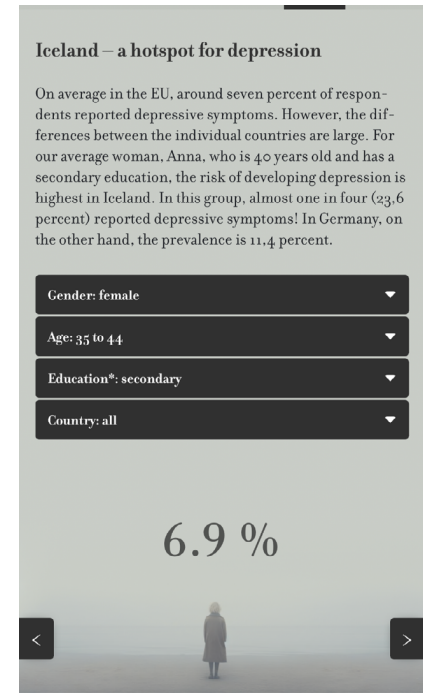

VISAP'25, Pictorials and annotated portfolios.

### Ethical Care:
### One person as a stand-in for a group

One challenge for designers of data visualizations in public science communication is to capture and sustain attention, particularly when dealing with statistical content. Research by Villanueva et al. suggests that conventional formats, like bar charts, are often perceived as less engaging by non-expert audiences [4]. Franconeri et al. further explore the interplay of engagement and information visualization tailoring [5]. In today's digital media landscape, visual representations of science must compete with the aesthetic norms of online platforms, where photography dominates and familiarity with images plays a key role in viewer engagement.

News media coverage of depression, for instance, frequently relies on photographic imagery depicting individuals in everyday settings, as Wang illustrates, providing a sense of intimacy and humanizing the otherwise abstract topic of depression [6]. These images typically portray one person as a representative of a broader group of those affected.

Generative AI introduces a novel possibility for visualizing data: the capacity to produce multiple, demographically varied portraits that represent different segments of a dataset. This strategy holds particular promise for enhancing perceived relevance in public science communication. When viewers recognize individuals who resemble themselves – whether in terms of age, gender, or other socio-demographic attributes – they may feel a connection to the data being presented. This idea is explored by Rodighiero and Cellard [7], who examine how self-recognition operates in the context of data visualization, particularly when individuals identify their own digital traces in visual representations. The phenomenon can also be further understood through the lens of Social Identity Theory, as developed by Tajfel and Turner [8], which posits that individuals construct their self-concept partly through identification with social groups. The distinction between in-groups and out-groups influences perception, empathy, and trust. By visualizing statistical data through representations reflecting diverse group identities, designers may foster stronger identification with the subject matter – potentially increasing the salience of the information.

One approach explored in the project involved generating portraits based on biometric proxy data (age, gender, and educational background). While this technique allows for demographic tailoring, it also reveals the persistent gap between statistical categories and lived experiences. For instance, not all women aged 45–54 with tertiary education look alike, nor can any single image truly represent every individual of a group. Nevertheless, such visual approximations may offer a way to make data feel more personally meaningful, while foregrounding the tension between representation and specificity in public science communication.

After the initial experiments, the team observed that images with more realistic, front-facing portraits, such as those seen in Fig. 8-11 on the next page, appeared too specific, suggesting individual stories rather than representing a collective. To emphasize that each image represented a broader demographic group instead of a specific individual, the team experimented with different compositional choices.

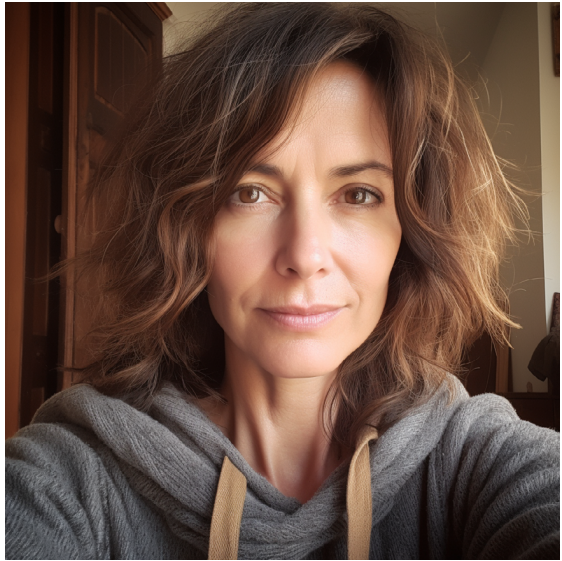
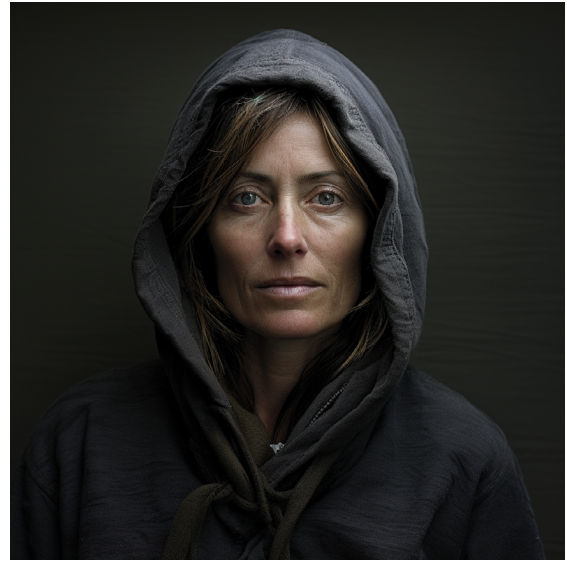

Fig. 8, prompt: selfie of a female, 45 to 54 years old, from Spain, primary education, melancholic atmosphere --v 5.2.

Fig. 9, prompt: an image of a woman, 45 to 54 years old, from Spain, in the style of haunting portraits, pigeoncore, minimalist images, atmospheric woodland imagery, realistic images, flickr, eerily realistic --v 5.2.

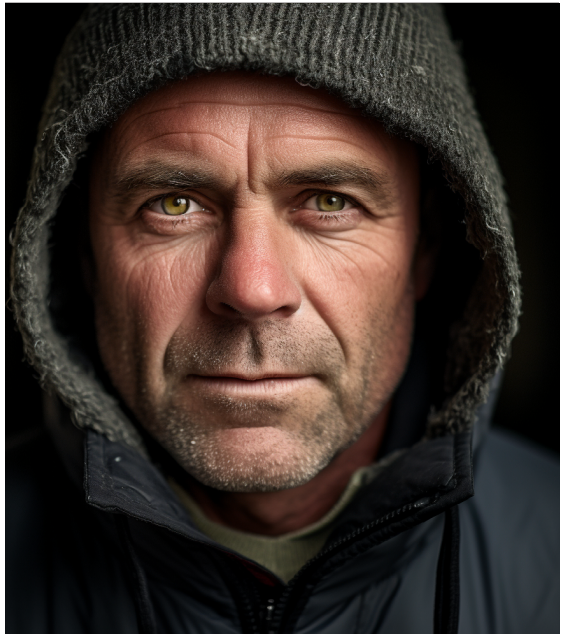
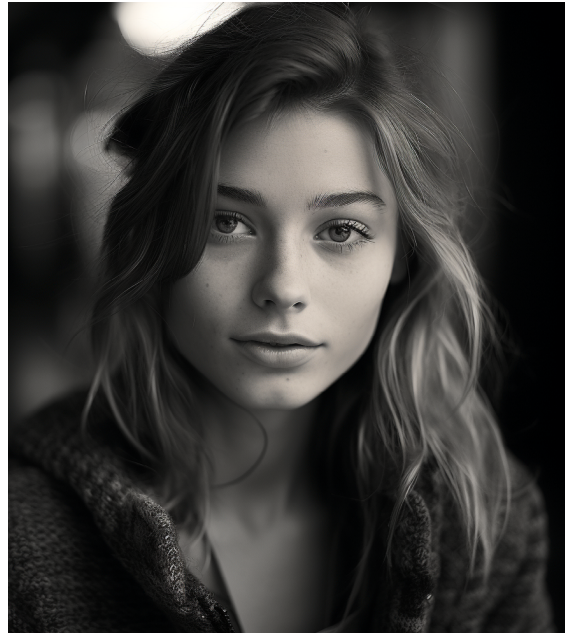

Fig. 10, prompt: a portrait of a male, 45 to 54 years old, from Luxembourg, photography, shallow depth of field, Leica --v 5.2.

Fig. 11, prompt: a portrait of a female, 15 to 24 years old, from Luxembourg, photography, shallow depth of field, Leica --v 5.2.

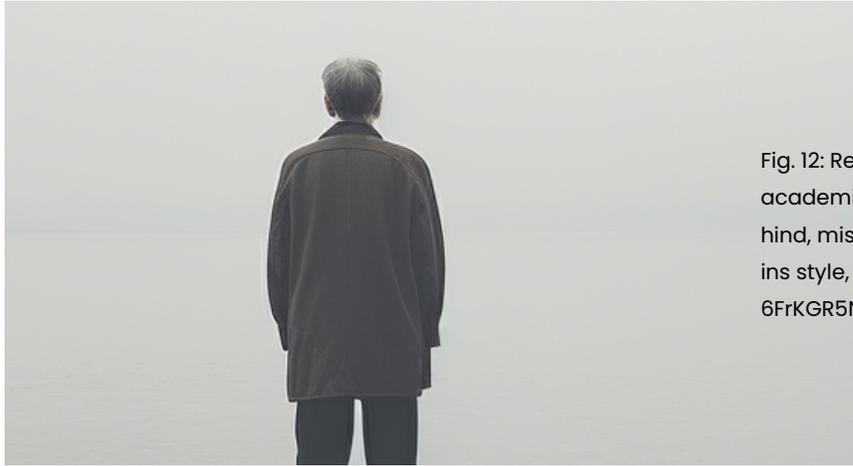

Fig. 12: Refined prompt: male, elderly, academic clothes, asian, from behind, mist, seascape, cinematic deakins style, wide shot --ar 21:9 --style 6FrKGR5N8TG4 --stylize 250 --v 5.2.

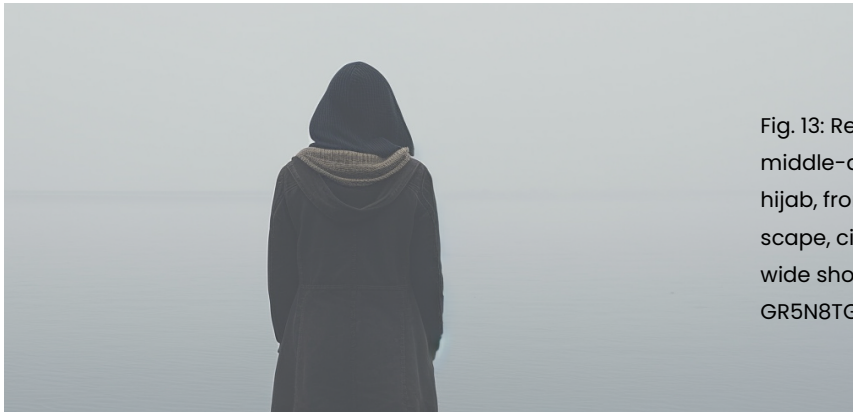

Fig. 13: Refined prompt: female, middle-aged, neutral clothes, hijab, from behind, mist, seascape, cinematic deakins style, wide shot --ar 21:9 --style 6FrKGR5N8TG4 --stylize 250 --v 5.2.

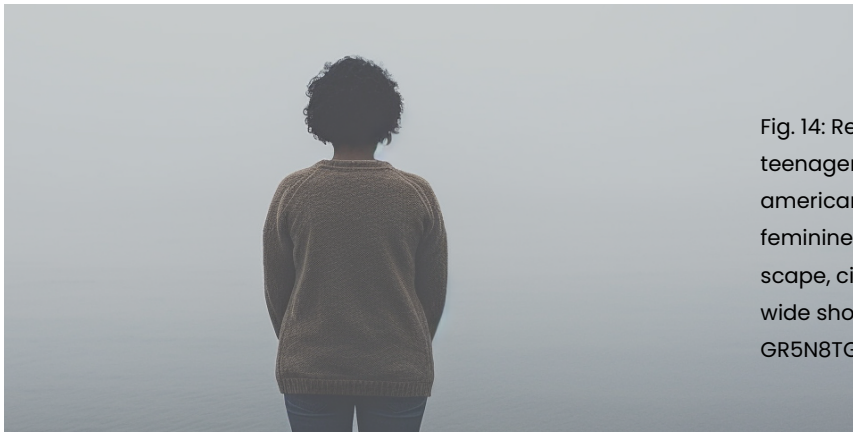

Fig. 14: Refined prompt: female, teenager, neutral clothes, african american, ultra short hair, dark skin, feminine, from behind, mist, seascape, cinematic deakins style, wide shot --ar 21:9 --style 6FrKGR5N8TG4 --stylize 250 --v 5.2.

## Technical Care: Prompt Engineering

Working with Midjourney 5.2 to visualize demographic diversity quickly revealed both technical and cultural limitations. Midjourney is a text-to-image model based on diffusion techniques that generate high-resolution, stylized images through iterative prompt engineering [9]. While the model reliably produced images of young, slim, light-skinned individuals – particularly women – generating more varied representations required extensive prompt refinement. Older individuals, people of color, and non-normative body types were often rendered inaccurately or not at all (see Fig. 12-14). For example, prompts such as "African" produced light-skinned figures unless supplemented with descriptors like "African American, ultra-short hair, dark skin, feminine" (see Fig. 14).

These challenges reflect not only in the architecture of the AI model, but also in the biases embedded in the vast corpus of training images – biases that mirror the visual norms of dominant online media, where people of color often remain underrepresented, as for example highlighted by Massie et al. [10]. This insight underscores an important point: the limitations of AI-generated representation are not solely a technical issue, but a reflection of broader cultural asymmetries in the visual data produced and circulated.

As images were generated in batches, the prompts – as illustrated in Fig. 12-15 – followed a consistent structure. Initially, gender was designated as a variable. Secondly, rather than utilizing numerical age values such as 15 to 14 years from the dataset, we employed descriptive language such as "teenager," as Midjourney responded more accurately to natural language formulations. Thirdly, educational attainment was encoded through clothing descriptors. The use of "casual" dress was employed to denote primary education, "mature" attire was used to signify secondary education, and "academic" clothing was used to represent tertiary education. A "neutral" style was employed in the absence of an educational filter. However, these choices were not entirely objective; the mapping of educational levels to clothing styles inevitably entailed the application of stereotypical associations. For instance, it was assumed that individuals with lower levels of formal education would dress more informally than those with higher academic attainment. During the prompt development phase, a paucity of diversity in the outputs was also observed. The phenomenon described above manifested in a manner that extended beyond the ethnic underrepresentation illustrated in Fig. 12-14. It also encompassed individuals of advanced age. The term "elderly" frequently failed to produce convincingly aged appearances. In an effort to elicit the generation of imagery depicting individuals of advanced age, we opted for a more precise approach by employing specific descriptors such as "octogenarian" and "nonagenarian." However, these initial prompts still proved to be inadequate. To enhance the effectiveness of the study prompts, we employed additional visual cues, such as "sitting on a bench, aged, diminutive, white hair, stooped, frail," with the objective of eliciting more precise representations of advanced age (see Fig. 15).

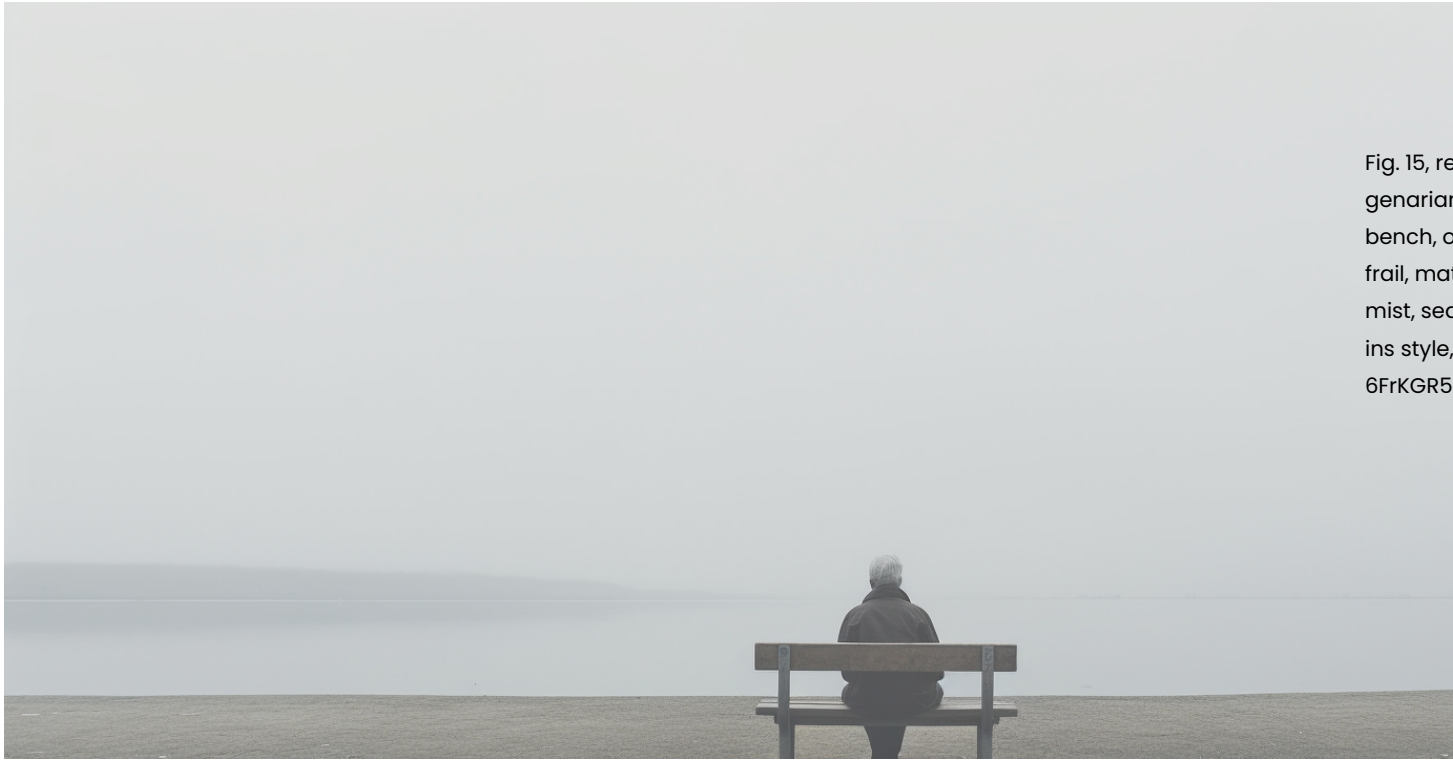

Fig. 15, refined prompt: female, octogenarian, nonagenarian, sitting on a bench, old, small, white hair, stooped, frail, mature clothes, from behind, mist, seascape, cinematic deakins style, wide shot --ar 21:9 --style 6FrKGR5N8TG4 --stylize 250 --v 5.2.

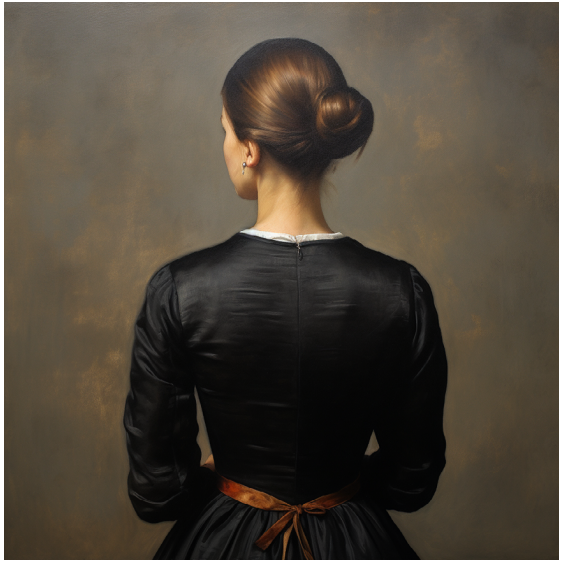
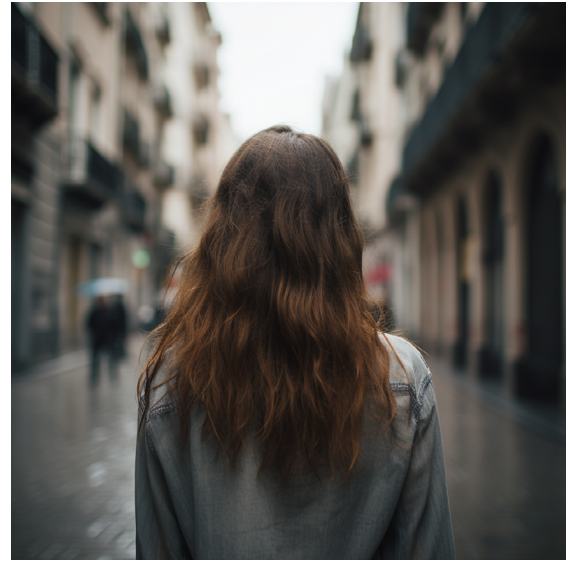

On the left Fig. 16, prompt: 17th century minimalistic Netherlandish oil painting of a female from behind, 45 to 54 years old, from Spain, primary education, melancholic atmosphere --v 5.2

On the right Fig. 17, prompt: a portrait of a female from behind, 15 to 24 years old, from Spain, melancholic atmosphere, shallow depth of field, Leica --v 5.2

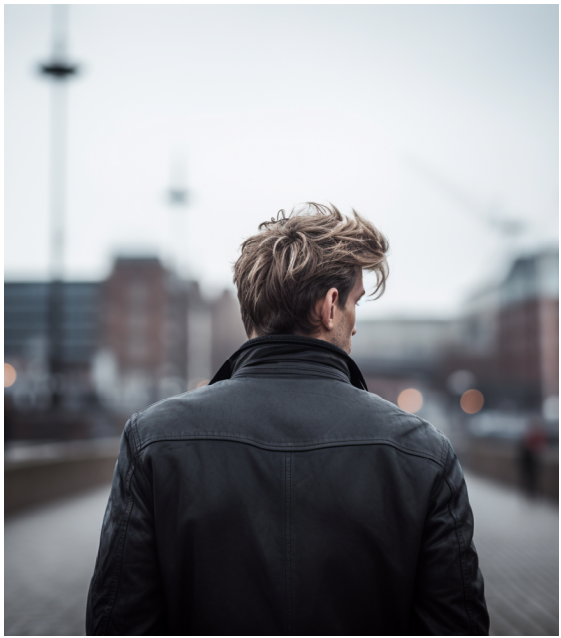
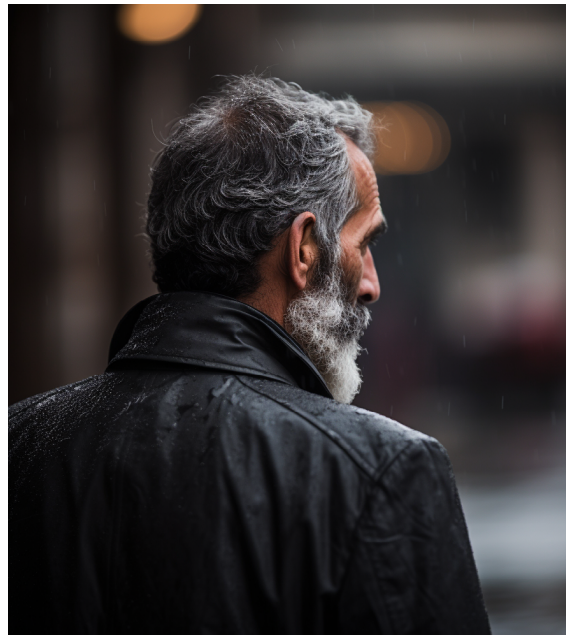

On the left Fig. 18, prompt: a portrait of a male from behind, 25 to 34 years old, from Netherlands, melancholic atmosphere, shallow depth of field, Leica --v 5.2

On the left Fig. 19, prompt: a portrait of a male from behind, 45 to 54 years old, from Turkey, melancholic atmosphere, shallow depth of field, Leica --v 5.2

The last part of the prompt consisted of a description of scenery and athomsphere. For this part a key design shift emerged with the use of a compositional strategy drawn from Caspar David Friedrich's Rückenfigur – a motif in which the subject is shown from behind. This visual distance encouraged viewers to see the figure not as a specific individual, but as a representative of a broader group. Informed by visual narrative theory and cinematic conventions, this perspective invited a more projective mode of engagement [11].

To shape the atmosphere of the images, we aligned the visual concept with the metaphor of depression as a confrontation with the unknown, which developed to be represented by a vague, mist-laden landscape (see experiments in image 20 and 21). We drew inspiration from the work of cinematographer Roger Deakins, whose visual style we explicitly referenced in the image prompts, as can be seen in the Fig. 22-26. Deakins is known for using lighting and environment to evoke internal struggle without prescribing fixed emotional states, as analyzed by Navas and Rubio [12]. After experimentation with stylistic modifiers, the use of "cinematic Deakins style" produced an evocative yet restrained tone that harmonized with the subject matter.
However, this prompt choice was not unproblematic: referencing a visual style shaped by a white, male cinematographer within Western mainstream cinema may risk reproducing aesthetic biases encoded in that tradition through image generation. As with all design choices, prompt engineering involves visual assumptions that merit critical scrutiny.

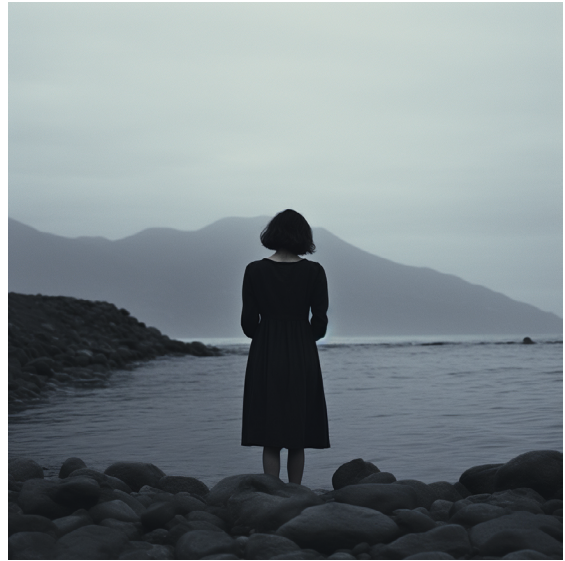

To ensure stylistic consistency across the dataset, the refined prompt served to train Midjourney's Style Tuner (version 5). From 128 generated style variations, the team selected the ones most aligned with the communicative goals of the project.

Top, Fig. 20, prompt variation: melancholic atmosphere, foggy evening, wet plate negative.

Bottom Fig. 21, prompt variation: in the style of dark, moody landscapes, minimalist landscapes, gabriel isak, eerily realistic, animated gifs, emotionally charged portraits, loose and fluid.

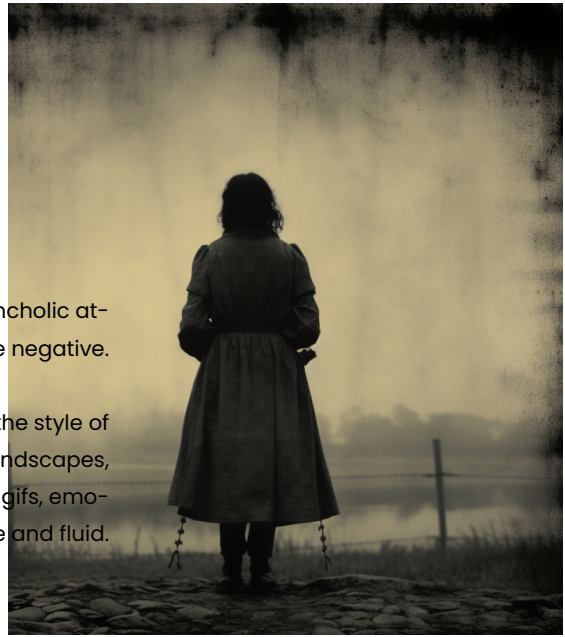

Although the project did not explicitly aim to critique Western visual norms, we acknowledge in this reflection on our visual style that our design decisions have certainly been informed by our backgrounds as designers and researchers shaped by European contexts, from prompt formulation to image selection.

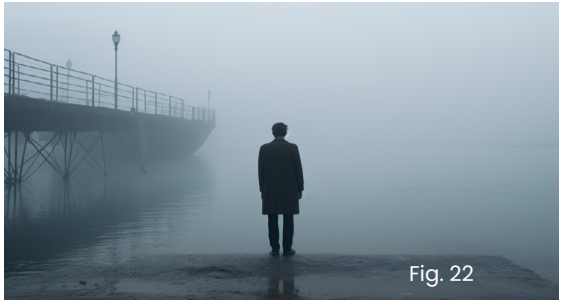

Fig. 22

Prompt variations Fig. 22 & 23: the solo woman/man standing in the fog, looking around, in the style of minimalist starkness, muted seascapes, roger deakins, dark white and light blue, introspective.

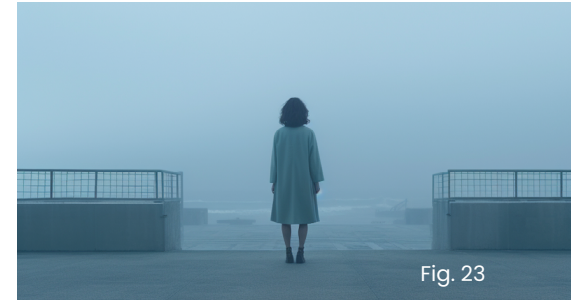

Fig. 23

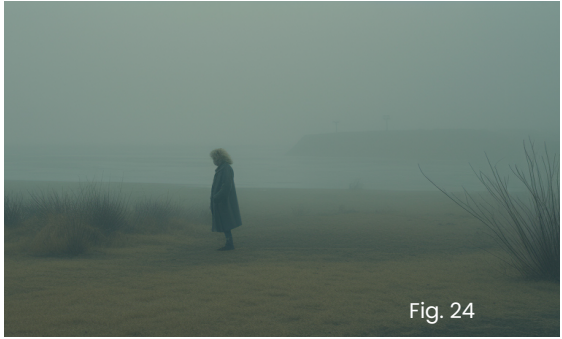

Fig. 24

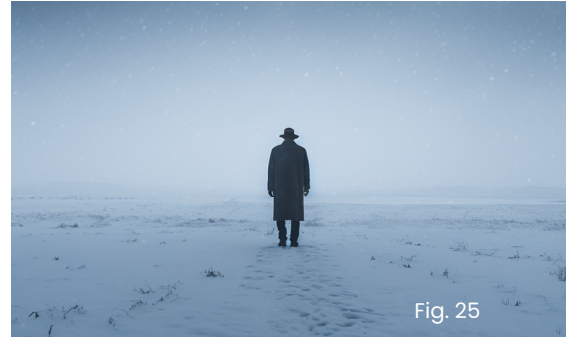

Fig. 25

Fig. 24 prompt: the solo woman, 45 to 54 years old, from Spain, standing in the fog, looking around, in the style of minimalist starkness, muted seascapes, roger deakins.

Fig. 25 & 26 prompt: the solo man standing in the snow, looking around, in the style of minimalist starkness, jessica drossin, muted seascapes, roger deakins, dark.

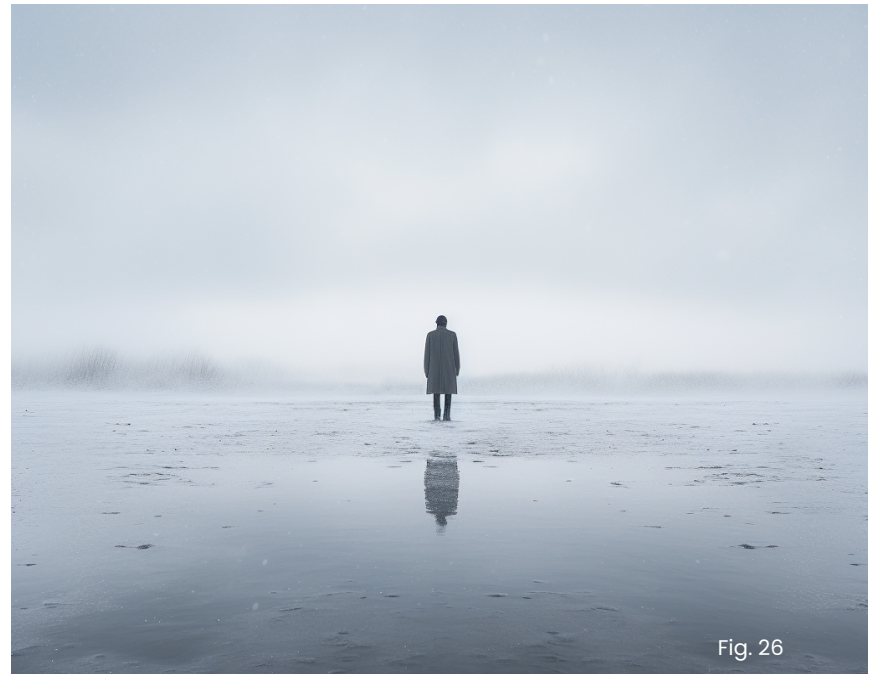

Fig. 26

## Visualizing the Topic of Depression – Between Atmosphere and Cliché

In developing the background imagery for the interactive bar chart, the team faced a dual challenge: how to visually evoke the topic of depression without falling into reductive motifs, and how to design within the technical constraints of a dynamic, horizontally oriented layout. The composition needed to be visually calm and structurally unobtrusive, allowing space for the chart itself while still pointing toward the theme. While the chosen aesthetic aligned with our aim to convey introspection and ambiguity, it also echoed a familiar visual language often used to depict depression: grey tones, solitude, and a sense of melancholy. These stylistic choices, while legible and emotionally restrained, risked reinforcing dominant visual clichés which may overlook the full spectrum of depressive experience.

This tension highlights a central concern of designing with visual care: how to offer visual cues without reducing complex, socially stigmatized conditions to static symbols. In future iterations, a key question will be how to balance atmosphere with visibility – ensuring the image is not so subtle that users overlook it, while also avoiding excessive direction or distraction. Audience feedback from early presentations suggests that the image changes were often too inconspicuous to register. Understanding how image salience interacts with attention, perception, and interpretation remains an open question for further investigation.

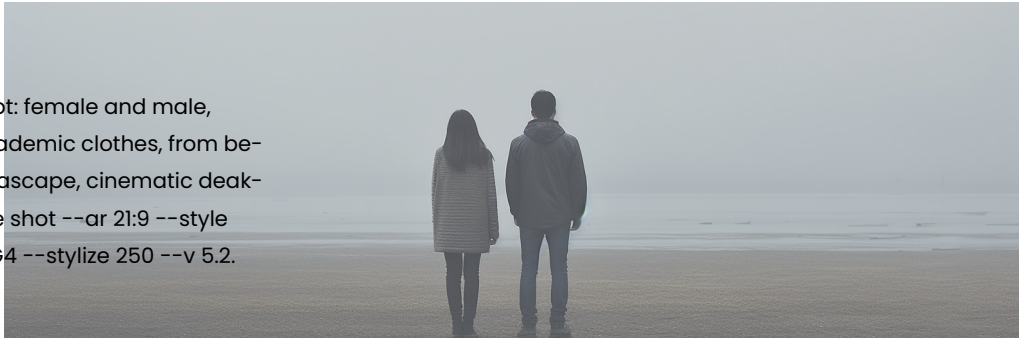

Fig. 27, prompt: female and male, teenager, academic clothes, from behind, mist, seascape, cinematic deakins style, wide shot --ar 21:9 --style 6FrKGR5N8TG4 --stylize 250 --v 5.2.

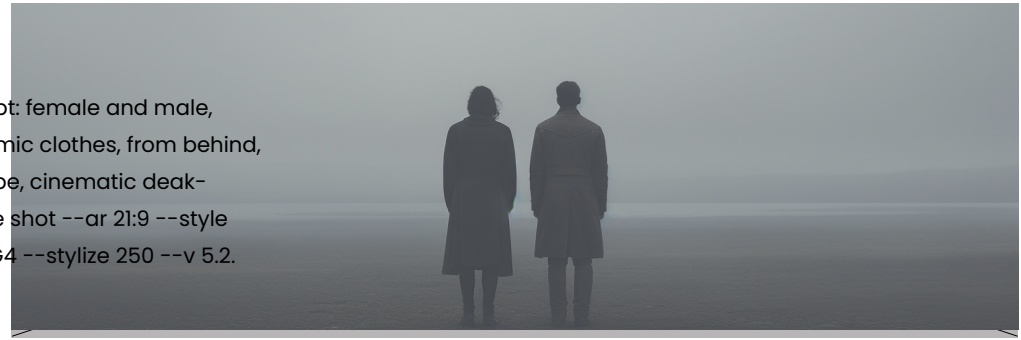

Fig. 28, prompt: female and male, adult, academic clothes, from behind, mist, seascape, cinematic deakins style, wide shot --ar 21:9 --style 6FrKGR5N8TG4 --stylize 250 --v 5.2.

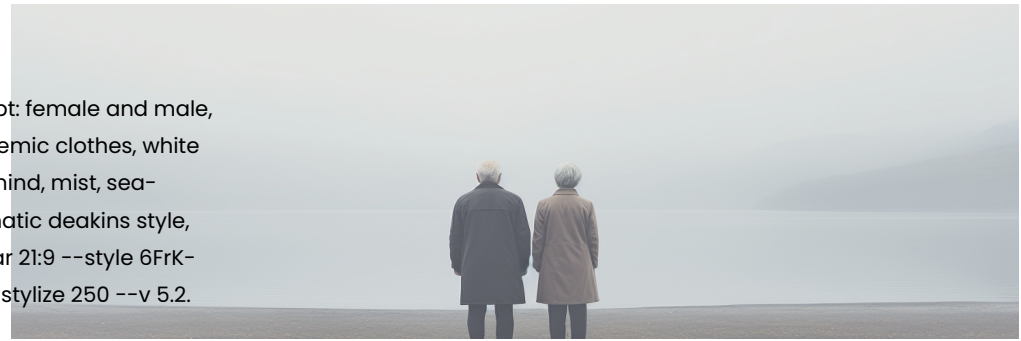

Fig. 29, prompt: female and male, elderly, academic clothes, white hair, from behind, mist, seascape, cinematic deakins style, wide shot --ar 21:9 --style 6FrKGR5N8TG4 --stylize 250 --v 5.2.

## Participatory Workshop: Prompting Depression, Together

To address stereotypical visual imagery associated with depression, a participatory workshop was organized. While the invitation was extended to all Spektrum der Wissenschaft subscribers, the workshop was attended by 11 women, mostly over 40 with academic background. Participants were invited to generate their own Midjourney prompts and discuss the visual language surrounding depression. The workshop revealed a broader spectrum of associations: several participants created bright, open, or emotionally ambivalent scenes (see e.g. Fig. 30, 34, 35), challenging the project's initial low-key aesthetic. These co-created images emphasized that depression has many faces, some visible and others more subtle. Participants voiced concern that consistent use of dark imagery may reinforce narrow, stereotypical conceptions of mental health. At the same time, discussions emerged about the risk of down playing the condition: images perceived as overly light-hearted or serene – such as the cheerful breakfast scene seen below in Fig. 30 – were considered potentially misleading or dismissive of the severity of depression.
In response, the designer refined the visual concept through iteration, experimenting with new stylistic directions. The resulting postcard series (see Fig. 36 -38), displayed at a follow-up event, allowed for informal feedback: the watercolor-style version was most frequently taken by visitors, indicating broader resonance and a willingness to more diverse visual interpretations.

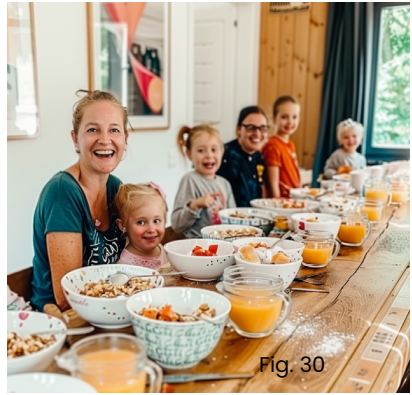
Fig. 30

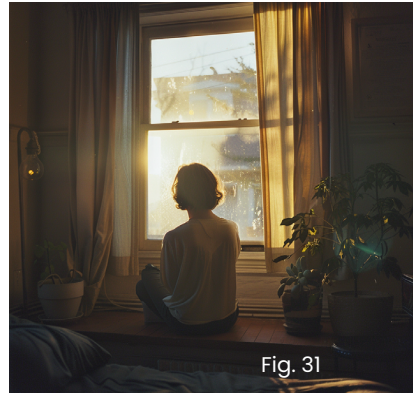
Fig. 31

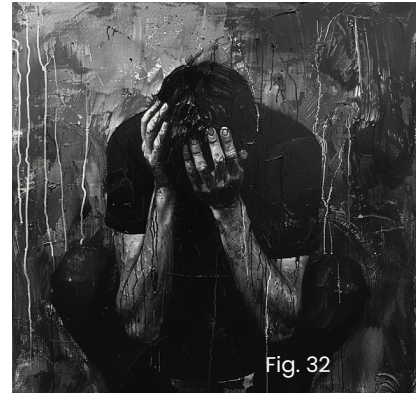
Fig. 32

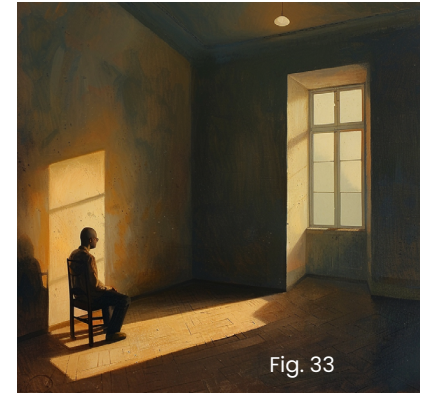
Fig. 33

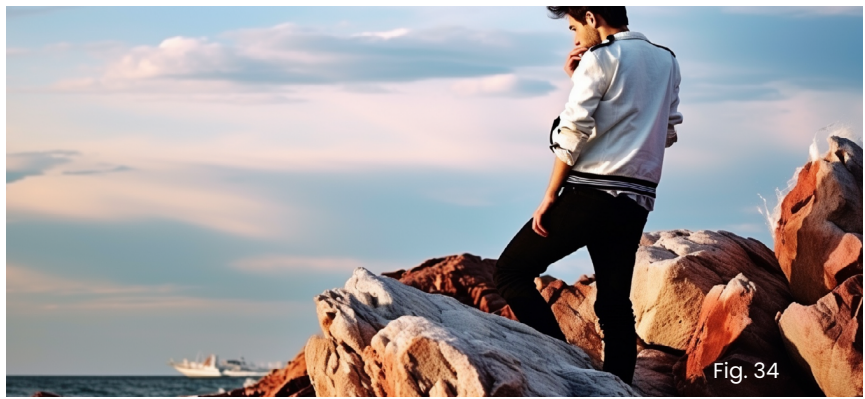
Fig. 34

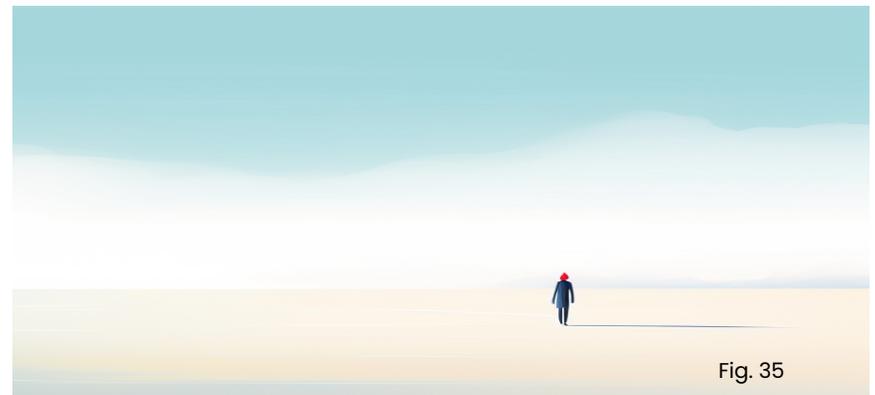
Fig. 35

## Conclusion – Care as Visual Method

This project set out to explore how generative AI might help audiences see "the people behind the data" – but along the way, it revealed how easily technical convenience can reinforce normative visualities. What began as an attempt to humanize statistical prevalence data evolved into a deeper investigation of what it means to design with care in the stigmatized and socially sensitive context of depression.

Throughout the process, it became clear that neither generative models nor their outputs are neutral. From prompt design to aesthetic framing, every decision shapes how mental health is represented – or neglected. Our workshop findings highlighted both the limits of our initial visual language and the potential for more diverse and resonant interpretations. In this sense, care is not a static value but a method: iterative, dialogic, and open to critique.

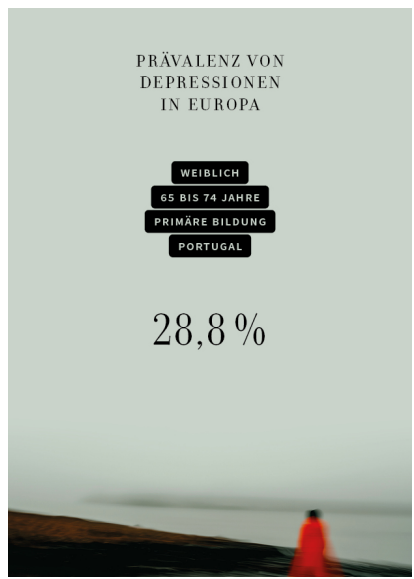

Fig. 36.

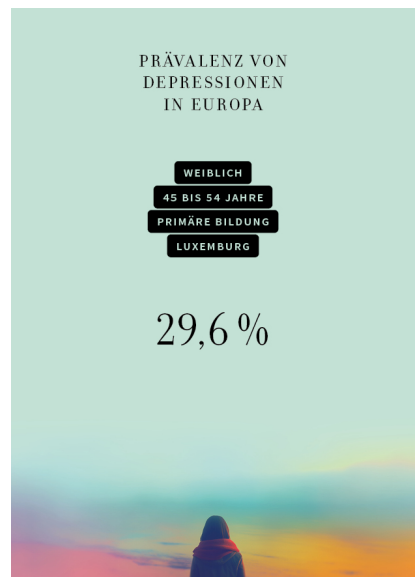

Fig. 37.

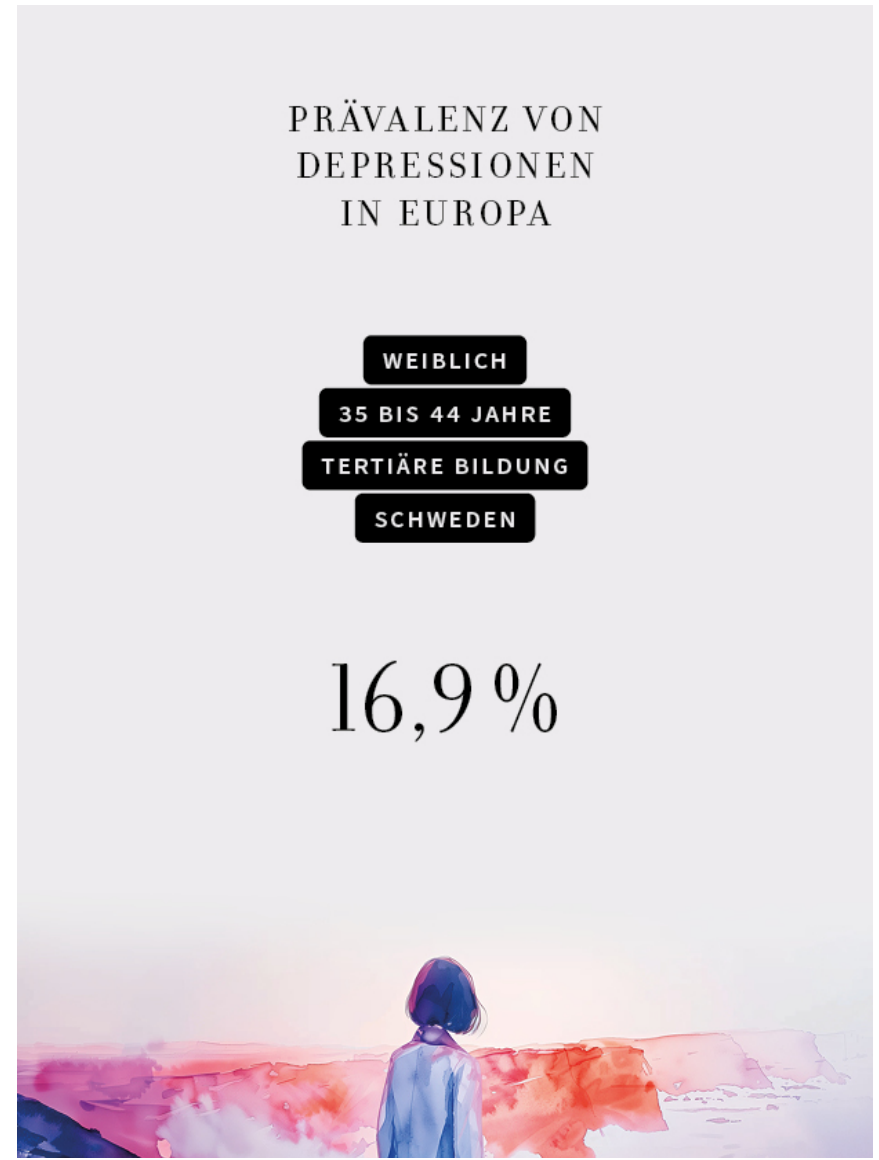

Fig. 38.

As science communicators and designers working with AI, we must be vigilant not only about the images we generate but also the assumptions we carry. Navigating the line between care and bias means acknowledging that representation always involves a choice – and that thoughtful, participatory design can challenge default modes of seeing.

## Acknowledgements
The KielSCN communication designer Björn Döge led the visual and conceptual development as well as the implementation of the project via observable plot [13]. This project was developed within the Kiel Science Communication Network, a transdisciplinary research initiative focused on advancing visual strategies in science communication.